# PROTOSTAR MASS DUE TO INFALL AND DISPERSAL

Philip C. Myers

Harvard-Smithsonian Center for Astrophysics, 60 Garden Street, Cambridge MA 02138 USA

pmyers@cfa.harvard.edu

## ABSTRACT

The mass of a protostar is calculated from the infall and dispersal of an isothermal sphere in a uniform background. For high contrast between peak and background densities and for short dispersal time $t_d$, the accretion is "self-limiting": gas beyond the core is dispersed before it accretes, and the protostar mass approaches a time-independent value of low mass. For lower density contrast and longer dispersal time, the accretion "runs away": gas accretes from beyond the core, and the protostar mass approaches massive star values. The final protostar mass is approximately the initial gas mass whose free-fall time equals $t_d$. This mass matches the peak of the IMF for gas temperature 10 K, peak and background densities $10^6$ and $10^3$ cm$^{-3}$, and $t_d$ comparable to the core free-fall time $t_{core}$. The accretion luminosity exceeds 1 $L_O$ for 0.1 Myr, as in the "Class 0" phase. For $t_d/t_{core}$=0.4-0.8 and temperature 7-50 K, self-limiting protostar masses are 0.08-5 $M_O$. These protostar and core masses have ratio 0.4 ± 0.2, as expected if the core mass distribution and the IMF have the same shape.

*Subject headings:* ISM: clouds — ISM: jets and outflows — stars: formation

# 1. INTRODUCTION

The origin of stellar masses is a key problem in astrophysics. The course of stellar evolution and many stellar properties are tied to the value of the stellar mass. Understanding star formation requires understanding of how stars form with a mass range of order $10^3$. The initial mass function of stars (IMF), or the distribution of stellar masses at birth, is a property which appears similar, if not identical, over a wide range of settings in the Milky Way and in other galaxies (Chabrier 2005). Yet the origin of stellar masses is poorly understood, and most available models are either incomplete or in conflict with observations.

## 1.1. Star formation models

It is well understood how self-gravity can concentrate gas in the presence of magnetic fields, turbulence, rotation, and thermal pressure, leading to protostar formation and accretion (McKee & Ostriker 2007, Adams & Shu 2007, Ballesteros-Paredes et al 2007). Yet such descriptions of the star formation process are incomplete unless they also explain how protostar accretion decreases with time to yield a well-defined protostar mass.

Collapse of a bounded initial condensation yields a fixed protostar mass by definition. Such collapses have been studied in detail, for a uniform self-gravitating sphere with a static outer boundary (Larson 1969), and for a centrally condensed "Bonnor-Ebert" sphere (Bonnor 1956, Ebert 1955, Foster & Chevalier 1993, Ogino, Tomisaka, & Nakamura 1999). Each of these model initial states has an infinitely steep transition to a very low-density environment, which is not observed in star-forming clouds.

Collapse of an unbounded initial condensation gives a well-defined protostar mass, independent of maximum radius, provided its density is nonsingular at the origin and declines with radius as $r^{-p}$, with $p > 3$. Yet most studies of cores and their environments indicate $p \approx 2$ in cores and $p < 2$ around cores (Bergin & Tafalla 2007). Even B335, the prototype of an isolated star-forming globule, is surrounded by extended molecular gas of density $\sim 10^3$ cm$^{-3}$ (Frerking, Langer & Wilson 1987).

Magnetic tension forces may prevent the infall of a magnetically subcritical envelope around a supercritical core (Mouschovias 1976; Shu, Li & Allen 2004, hereafter SLA). This problem has been addressed for a singular isothermal sphere (SIS) threaded by an initially uniform, vertical, frozen-in field. The resulting core mass scale is similar to that of the critically stable BE sphere, with the external pressure replaced by the magnetostatic pressure. However, such static levitation requires unrealistically large field strength at the stellar surface. Instead, SLA suggest that the protostar mass is limited to its mass scale by winds which originate from a combination of weaker fields and disk rotation.

Ejection of a protostar from its parent core to a lower-density environment can yield masses of brown dwarfs and very low-mass stars (Reipurth & Clarke 2001). However, brown dwarfs are observed to have a fraction of circumstellar disks similar to that of ordinary low-mass stars, and the substellar and stellar populations show no significant difference in spatial distribution. It appears that the low masses of brown dwarfs do not generally arise from early ejection at high speed (Luhman et al 2007).

In competitive accretion models, moving protostars accrete different shares of the available gas. Simulations yield well-defined masses having many low values and a few high values, resembling the mass distribution of the IMF (Bonnell et al 1997). This picture assumes that the initial protostars have no surrounding cores. Yet protostars in nearby star-forming regions are highly correlated with dense cores (Enoch et al 2008, Jørgensen et al 2007, 2008), and a young star gains significant mass during its "Class 0" phase, when it is still embedded in its parent core (White et al 2007). The close encounters expected in some models of competitive accretion are probably limited to regions of unusually high stellar density (Adams et al 2006).

1.2. Gas dispersal and outflows

None of the foregoing models includes the reduction of gas density available for accretion, due to dispersal of the accreting gas.

Gas disperses on the pc scale in young clusters, as indicated by the decrease in gas surface density by an order of magnitude, from a few × $10^{22}$ cm$^{-2}$ where protostars are forming, to a few × $10^{21}$ cm$^{-2}$ where young stellar objects become optically visible (Allen et al 2007). This dispersal removes star-forming gas from a cluster, and in most cases unbinds its stars (Lada & Lada 2003). The most powerful agents of dispersal are ionization and winds from OB stars. Clusters which lack OB stars also disperse much of their gas within 1 Myr, likely due to winds and radiation from their lower-mass stars. Such dispersal removes low-density cluster gas and also erodes or evaporates its dense cores. Well-studied examples of dispersing cores include the bright-rimmed globule TC2 in the Trifid Nebula (Lefloch et al 2002) and the Finger globule in the Carina Nebula (Smith, Barbá & Walborn 2004).

In nearby star-forming regions, statistics of association between dense cores and young stellar objects indicate dispersal of dense cores and their environs in significantly less than 1 Myr. Among 265 Spitzer sources in Ophiuchus and 353 in Perseus, all Class 0 protostars have associated SCUBA cores, half of the Class I protostars have such associated cores, but fewer than 3% of the Class II and Class III sources have such associated cores (Jørgensen et al 2008). Thus protostars lose their cores at an evolutionary age typical of Class I, and much less than that of Class II. A recent estimate of the Class I duration is 0.4-0.5 Myr, based on observations of 1035 young stellar objects in the five nearby "c2d" clouds (Evans et al 2008). For constant birthrate, the average age of a Class I YSO is equal to half the Class I duration, or 0.2-0.3 Myr.

These statistics indicate that once the typical core forms a protostar, the core becomes undetectable within a few 0.1 Myr. For isolated cores, the most likely agent of dispersal is the molecular outflow driven by winds from the protostar-disk system (Snell, Loren & Plambeck 1980, Arce et al 2007). In groups and clusters, core gas and its surrounding gas may also be dispersed by external heating, ionization, and winds from neighboring stars, as described above. Young stars could in principle separate from their parent cores by migration, but imaging and kinematic studies indicate that such migration speeds are limited to values $< 0.1$ km s$^{-1}$, too low to be significant (Walsh et al 2006, Jørgensen et al 2007).

The opening angles of protostellar outflows tend to increase with increasing age of the protostar (Velusamy & Langer 1998, Arce & Sargent 2006, hereafter AS), and this property indicates reduction of the gas available for infall onto low-mass stars (Shu, Adams & Lizano 1987, Velusamy & Langer 1998, AS) and onto Herbig Ae and Be stars (Fuente et al 2002). Gas dispersal during accretion has been described by spherical outflows (Shu et al 1991, Nakano, Hasegawa & Norman 1995, Adams & Fatuzzo 1996) and by outflows with fixed collimation (Matzner & McKee 2000). The response of a medium to a widening conical jet of fixed radial velocity has been described by Cantó, Raga & Williams (2008).

1.3. Overview of paper

In this paper, protostar masses are predicted for an initial system which is more extended and more realistic than a BE sphere, and whose accretion as a function of time is the net result of infall and dispersal. Sections 2, 3, and 4 describe respectively the initial conditions, the dispersal and infall model, and the results of the calculations. Section 5 discusses uncertainties and implications and Section 6 summarizes the paper. The main result is that a core with sufficient density contrast can form a protostar with well-defined mass if the time scale of the gas dispersal is comparable to the core free-fall time.

2. INITIAL CONDITIONS

2.1. Introduction

The initial gas configuration adopted here has an uniform inner region of high density and a uniform outer region of low density, separated by a region of sharply declining density. This structure is similar to starless core models based on absorption observations in the mid-infrared (Bacmann et al 2000), and on emission observations in millimeter lines and continuum (Tafalla et al 2006). The configuration is modelled as an isothermal sphere in equilibrium between self-gravity and thermal pressure ("IS"), superposed on a uniform background, and is denoted "IS+ U."

In these models the inner region has nearly constant density with central value $n_0$. This "flat-top" density profile is a characteristic feature of well-studied starless cores (Ward-Thompson et al 1994, Di Francesco et al 2007). The outer region has nearly constant density, approaching $n_u$, with $n_u \ll n_0$. This outer region represents the "clump" of lower-density gas on which the core is typically superposed, seen in visual or near-infrared extinction or in millimeter line emission (Bergin & Tafalla 2007). At intermediate scales the density declines with increasing radius r approximately as $r^{-2}$, as indicated by fits of IS models to dust emission and absorption observations of starless cores (Alves, Lada & Lada 2001, Evans et al 2001, Tafalla et al 2006).

The uniform background is the simplest possible nonzero background, and is a first approximation to the core environment. For backgrounds having nonuniform structure and nonspherical geometry, the main change is to reduce the importance of runaway accretion, as discussed in Section 5.

The IS+U model is more realistic than the BE model for most regions of low-mass star formation. The IS+U density decreases smoothly from the core to the background, while the BE density decreases discontinuously across a sharp boundary. The outer IS+U density matches the typical density of observed molecular cloud clumps, while the outer BE gas more nearly matches the density and temperature in H II regions, which do not surround well-studied starless cores. Thus the IS+U model has gas available for accretion from beyond the BE radius, while the BE model does not.

The IS+U configuration is a slowly collapsing system which is close to equilibrium. Its dynamical status is a small departure from the case of zero background, when the system is in exact equilibrium. As the background density increases from zero, the mass within each spherical shell increases, with no change in pressure gradient. Thus the system becomes overdense and starts to collapse. This situation is similar to that in some numerical studies, where an equilibrium system is made overdense in order to initiate its collapse (e.g. Ogino, Tomisaka, & Nakamura 1999).

Although the IS+U system is initially collapsing, its initial speeds are negligibly small. They are less than the 0.05-0.1 km s$^{-1}$ inward speed reported from some starless cores (e.g. Lee, Myers & Tafalla 1999). Such observed motions may represent an initial condition for collapse, as discussed by Fatuzzo, Adams & Myers (2004, hereafter FAM) and as predicted by the ambipolar diffusion model of Adams & Shu (2007). On the other hand these same motions are also expected to arise during the early stages of collapse of a centrally condensed core initially at rest (Myers 2005, Kandori et al 2005). This latter view is assumed in the infall calculation of Section 3, where the initial infall speed is zero. If instead the initial infall speed is ~ 0.1 km s$^{-1}$ the infall time is reduced by a typical factor ~2 (FAM).

Dense cores and their backgrounds have considerable observational justification, but their physical origin is still poorly understood (McKee & Ostriker 2007). It is expected that they condense as self-gravity and thermal physics become more important than the effects of turbulence and magnetic fields (Larson 2003). It is not settled whether this condensation is controlled more by the physics of turbulent flows (Mac Low & Klessen 1994), ambipolar diffusion (Adams & Shu 2007), or MHD waves (Van Loo, Falle & Hartquist 2006). Since cores are well-studied by observations, it is not essential to model their formation in order to describe their structure.

2.2. Standard parameter values

The adopted standard gas temperature is T=10 K, typical of core temperatures derived from NH$_3$ line observations (Jijina, Myers & Adams 1999).

The standard peak density is $n_0 = 1 \times 10^6$ cm$^{-3}$. Among 13 well-studied starless cores in nearby clouds, six cores are "most evolved" according to their CO depletion, enhanced deuteration, broad line widths, and infall asymmetry. All six have estimated central density > 0.5 × $10^6$ cm$^{-3}$ while the rest of the cores typically have lower central density (Crapsi et al 2005).

The standard background density is $n_u$ = 1 × $10^3$ cm$^{-3}$, as in the core background models for L1498 and L1517 (Tafalla et al 2006), and the Pipe nebula (Lada et al 2008). This value is ~4 times greater than the mean density above $A_V$ =2 in the Perseus complex (Evans et al 2008) and ~3 times less than the typical mean density of the "extinction cores" in the same complex (Kirk, Johnstone & Di Francesco 2006, hereafter KJD).

2.3. Initial density profile

With these justifications, the IS+U density $\rho$ is given by

$$\rho = \rho_{IS}(1-f) + \rho_0 f \qquad (1)$$

where f is the ratio of uniform background density $\rho_u$ to peak density $\rho_0$, or $10^{-3}$ for standard density values. The isothermal sphere density $\rho_{IS}$ is

$$\rho_{IS} = \rho_0 \left( \frac{C}{c^2 + \xi^2} - \frac{D}{d^2 + \xi^2} \right) \qquad (2)$$

using the analytic approximation to the isothermal function by Natarajan & Lynden-Bell (1997), with C=50, D=48, c=√10 and d=√12. The dimensionless radius $\xi$ is related to the radius a by

$$\xi \equiv \frac{a}{a_{th}} = \frac{a}{\sigma}\sqrt{4\pi G\rho_0} \qquad (3)$$

where $a_{th}$ is the thermal scale length, $\sigma$ is the one-dimensional isothermal velocity dispersion, and G is the gravitational constant. The initial radius is denoted $a$ to distinguish it from the radius r at later times, as in section 3.

The mass within radius a is obtained by integrating eq. (1),

$$M(<a) = M_0 m(<\xi) \qquad (4)$$

where the mass scale $M_0$ is the mass of a critically stable Bonnor-Ebert sphere of the same temperature and peak density,

$$M_0 \equiv \frac{b\sigma^3}{G^{3/2}\rho_0^{1/2}} \qquad (5),$$

with b=4.43 (Bonnor 1956). In eq. (4) the dimensionless mass enclosed by $\xi$ is given by

$$m(<\xi) \equiv \frac{\alpha\xi}{b\sqrt{\pi}} \qquad (6),$$

where $\alpha$ is the dimensionless ratio of enclosed mass to radius, a function of $\xi$ given in eq. (A1).

In eqs. (6) and (A1), $\alpha$ is defined so as to reveal the nearly linear dependence of $m(<\xi)$ on $\xi$. For sufficient contrast between peak and background densities, f < ~0.003, $\alpha$ is nearly constant over a significant range of ξ. In this range, the dependence of mass on radius is essentially linear, as it is for the SIS, but with a slightly greater coefficient. For $f=10^{-3}$, α lies within 1% of 1.30 for $8 < ξ < 29$. In the limiting case of the SIS, α=1 for all values of ξ.

The background gas available to accrete onto a core is assumed to extend in each direction until it meets the background gas associated with the neighboring core. Each such core-background unit is assumed to have condensed while conserving mass from the uniform Jeans mass $M_J$ of its background density,

$$M(<a_{max}) = M_J \equiv \frac{\pi^{3/2} \sigma^3}{G^{3/2} \rho_u^{1/2}} \qquad (7)$$

where $a_{max}$ is the outer radius and where the conventional definition of Jeans mass is used (Spitzer 1968).

For gas temperature 10 K and background density $10^3$ cm$^{-3}$, eq. (7) gives the Jeans mass as 17.6 $M_O$. The radius $r_J$ of a uniform sphere of the same mass is 0.42 pc. Due to the central concentration of the IS+U model, $a_{max}$ in eq. (7) is less than $r_J$, and depends on the value of f. For standard parameter values, $a_{max} = 106\, a_{th} = 0.86 r_J = 0.36$ pc. In the limit $f \ll 1$, eq. (7) gives $ξ_{max} = 3.39 f^{-1/2}$.

Eq. (7) assumes that the background gas has the same velocity dispersion as the thermal core gas in eq. (3), even though the gas around many cores has significant turbulent motions (Fuller & Myers 1992). However, when a self-gravitating region of turbulent gas fragments, it typically forms about as many condensations as the number of thermal Jeans masses it contains (Larson 2005). It is assumed that an IS+U unit has the mass of such a condensation, as given in eq. (7).

2.4 Dense core definition

It is useful to quantify the densest gas in the model, to represent an "initial condition" for star formation. Since the density varies continuously with radius, any such definition is somewhat arbitrary. To represent a meaningful initial condition the dense core mass should be comparable to that of typical low-mass stars, of order 0.3 $M_O$ at the peak of the IMF (Chabrier 2005, Kroupa 2002). Its mass should also be comparable to that derived from well-resolved observations in nearby star-forming regions, such as the 58 submm cores in the Perseus complex, whose median mass is 0.6 $M_O$ (KJD).

Such a dense core has less mass than the IS+U mass above the background. To match 0.6 $M_O$ by integrating the IS+U model down from its peak density requires a minimum density ~2 × $10^4$ $cm^{-3}$. This density is much greater than the background density in nearby star-forming regions, of order $10^3$ $cm^{-3}$, as discussed in Section 2.1 above (regions forming massive stars may have higher backgrounds, as discussed in Section 5). If a core were defined as the gas whose density or column density exceeds the background, its mass (several $M_O$) would be too great and its free-fall time (~ 0.5 Myr) would be too long for the core to be a useful initial state for low-mass star formation.

Instead, it is required here that the IS+U density exceed the geometric mean of its peak and background densities, $\rho > \rho_m \equiv (\rho_0 \rho_u)^{1/2}$. With this definition, core gas occupies the upper half of the range of log density, as shown in Figure 1. For standard parameter values the minimum core density is $3 \times 10^4$ cm$^{-3}$, yielding a core mass 0.68 M$_O$, comparable to the IMF mass and to the typical core mass cited above.

This core definition is based on a minimum density, as in observations of spectral lines having a critical density for excitation. The standard core radius and mass match typical values from high-resolution observations of cores in their mm and submm continuum emission, especially when these cores appear "clustered." Then the size of a core is set by its projected separation from neighboring cores rather than by the radius where it meets a background (Enoch et al 2006). In contrast, isolated cores defined by their extinction above a smooth background tend to have larger masses: the median mass of 159 such cores in the Pipe nebula is 2.8 M$_O$ (Alves, Lombardi & Lada 2007).

The initial core is both "starless" and "prestellar" in the terms defined by Di Francesco et al (2007) and Ward-Thompson et al (2007), since it contains no star and is self-gravitating. However it is not assumed that the mass of a core fixes the mass of the protostar which forms at the core center. As shown in Sections 3 and 4, this mass can be less than or greater than the core mass, depending on the competition between infall and dispersal.

2.5 Dense core radius and mass

The core outer radius is obtained from the defining condition $\rho = \rho_m$, combined with eqs. (1) and (2). This combination yields an equation quadratic in $\xi^2$, whose solution gives the core outer radius as

$$\xi_{core} = \frac{2^{1/2} \delta}{f^{1/4}} \qquad (8)$$

Here δ is a factor which approaches unity as f decreases, given in eq. (A2). For $f=10^{-3}$, eq. (A2) gives $\delta = 1.17$ and then eq. (8) gives $\xi_{core}=9.38$. For standard parameter values, the dense core radius is then $a_{core}=0.032$ pc. This radius agrees well with the median core radius observed in Perseus, 0.036 pc (KJD).

The core mass follows from eq. (6) with $\xi_{core}$ obtained from eq. (8), giving

$$M_{core} = \left(\frac{2}{\pi}\right)^{1/2} \frac{\alpha_{core}\delta\sigma^3}{G^{3/2}\rho_m^{1/2}} \qquad (9)$$

For standard parameter values, eq. (A1) gives $\alpha=1.30$, so in eq. (9) $\alpha_{core}\delta = 1.52$ and $M_{core}= 0.68$ $M_O$.

The core mass in eq. (9) exceeds the mass of a SIS having the same minumum density by the factor $\alpha_{core}\beta$. As expected, eq. (9) reduces exactly to the SIS mass in the limit where f goes to zero while $\rho_m$ stays finite. The adopted core definition has internal temperature and density profile similar to those of an unstable BE sphere, since $\xi_{core}=9.8$ for standard parameters, exceeding the critical BE value $\xi=6.46$ (Spitzer 1968). However as discussed above, the IS+U and BE environments differ significantly.

2.6 Bonnor-Ebert, core, and Jeans masses

Figures 1 and 2 show the dependence of density on radius and of enclosed mass on radius for the adopted IS+U model with the standard parameter values of Section 2.2. The radii increase from critical BE sphere to dense core to Jeans mass. In each figure the innermost and outermost regions are nearly uniform and have similar slope, while the core region of rapidly decreasing density has distinctly different slope. In Figure 1 the log density for the core boundary lies halfway between the peak and background values of log density, as expected from the core definition.

2.7  Mass and free fall time

The free fall time of the mass within initial radius a can be written

$$t_{ff}(a) \equiv t_0 \theta_{ff}(\xi) \qquad (10)$$

where $t_0$ is the free fall time of the peak density (Hunter 1962),

$$t_0 \equiv \left(\frac{3\pi}{32G\rho_0}\right)^{1/2} \qquad (11),$$

and where the dimensionless free fall time is

$$\theta_{ff}(\xi) \equiv \left[\frac{\xi^3}{m(<\xi)}\right]^{1/2} \qquad (12).$$

Then using eq. (6) the free-fall time of the gas within $\xi$ can be written

$$t_{ff} = \frac{t_0 \xi}{(6\alpha)^{1/2}} \quad (13)$$

and the gas mass within $\xi$ is given in terms of its corresponding free-fall time by

$$M(<\xi) = \frac{8\alpha^{3/2} \sigma^3 t_{ff}}{\pi G} \quad (14).$$

Application of the dense core radius obtained from eq. (8) to eq. (13) yields a dense core free-fall time of 0.12 Myr for standard parameter values.

### 3. GAS DISPERSAL AND INFALL

3.1. Dispersal

CO maps indicate that the typical molecular outflow extends well beyond the dense core, into its surrounding gas (Arce et al 2007). In the largest catalog of CO outflows, the mean diameter along the major axis is 0.7 pc for 129 outflows and the mean maximum speed is 10.7 km s$^{-1}$ for 217 outflows (Wu, Huang & He 1996). The typical outflow radius is therefore similar to the radius of the IS+U unit, 0.3 pc for standard parameters. A narrow jet rapidly punches through the dense gas with speed ~100 km s$^{-1}$ (Ray et al 2007). Then the molecular outflow removes core gas with maximum speed ~10 km s$^{-1}$, through a cavity whose width increases over time. When the cavity shape is well-defined, it is roughly conical or parabolic (AS).

Thus it is assumed here that the initial dispersal of the core and its associated background gas occur primarily through the widening of a conical outflow cavity, as a function of the time t since the start of infall, over a characteristic time scale $t_d$.

The mass of a shell between radii r and r + dr at time t is then

$$dM(r, t) = dM(a)\nu(t) \quad (15),$$

where the shell radius is denoted by a at the initial instant and by r at later times (Hunter 1962), and where $\nu(t)$ is the fraction of IS+U mass which has survived dispersal. The initial shell mass dM(a) is

$$dM(a) = 4\pi a^2 da \rho(a) \quad (16),$$

where $\rho(a)$ is the initial density at a.

The dispersal time scale $t_d$ is a free parameter and is not tied to the stellar mass or to an infall time scale. Many outflow models tap the gravitational potential energy of the infalling system, and some obtain a fixed ratio between the stellar mass accretion rate and the mass outflow rate (Tomisaka 1998, Shu et al 1991, Behrend & Maeder 2001). Such a fixed ratio of rates is not assumed here, since it would preclude the possibility that gas dispersal terminates the infall. Section 4 discusses observational constraints on $t_d$ in relation to the core free-fall time.

The widening of an outflow cavity with time is described by a study of 17 CO outflows (AS). AS find that half-opening angles increase up to ~ 50 deg by ages ~ 0.01-0.1 Myr, followed by an increase with shallower slope for ages from 0.01-0.1 to 1 Myr. The same sources show a distinct decrease in $C^{18}O$ envelope mass with age, suggesting that the outflow acts to clear the circumstellar mass as its opening angle increases. Similarly, a deep optical image of the outflow in TMC-1 shows a conical outflow cavity with half-opening angle ~40 deg at an estimated age 0.1 Myr (Terebey et al 2006), and *Spitzer Space Telescope* observations of the outflow from the protostar L1527IRS show a biconical cavity with somewhat larger angles (Tobin et al 2008).

These observations suggest that the solid angle of a bipolar outflow cavity increases from an initial value of zero, through a rapid rise, followed by a slower increase toward its maximum of $4\pi$. Such a picture has been discussed in terms of models of "wide-angle winds" or combinations of jets and wide-angle winds (Shu et al 1991, Shang et al 2006).

On the other hand, the detailed model of the B5 outflow cavity by Cantó, Raga & Williams (2008) widens with increasing radial distance from the protostar only at small radius, and then tapers to a narrow jet at radii greater than ~ 0.1 pc. Cavities having this shape, or having a parabolic shape, seem unable to remove all the circumstellar gas unless other agents of dispersal are also operating.

While outflows seem most important for initiating dispersal, they are likely to be assisted by erosion due to external outflows and ionization, as described in Section 1.2. Furthermore, once a significant mass fraction is dispersed by these means, the escape speed is reduced, unbinding some of the remaining gas. An isolated system in virial equilibrium becomes unbound if half its initial mass is removed rapidly (Hills 1980). For a biconical outflow in an initially spherical core, this condition corresponds to a half-opening angle of 60 deg. Core gas with such a wide cavity angle is also subject to increased heating through the cavity walls, promoting further unbinding. Thus the fast dispersal of gas by outflows may initiate a slow dispersal of remnant gas by gravitational unbinding.

The relative contributions of outflow, erosion, and unbinding to the dispersal of a core and its environs have not been studied in detail. In a simple description of the entire process, the cavity solid angle increases linearly with time and then more slowly,

$$\Omega_{out} = 4\pi\left[1 - \exp(-t/t_d)\right] \qquad (17)$$

where the value $t_d$ is a free parameter near 0.1 Myr. If this solid angle encloses negligible mass compared to the undisturbed gas, then the infalling gas at time t has surviving mass fraction

$$v(t) = 1 - \frac{\Omega_{out}}{4\pi} = \exp(-t/t_d) \qquad (18).$$

Alternate formulations of eqs. (17) and (18) are discussed in Section 5.

As a biconical outflow widens in an initially spherical collapsing core, the infalling gas departs from spherical symmetry, since the mass of each infalling shell occupies a solid angle which decreases from its initial value of $4\pi$. The resulting collapse concentrates some mass into a disklike structure, similar to the collapse of a slowly rotating sphere (Terebey, Shu & Cassen 1984, Boss 1987). In the following calculation such departures from spherical symmetry are ignored, since they are relatively small and since the goal is to obtain the total accreted mass as a function of time. Therefore each incomplete infalling shell is replaced by a complete infalling shell of the same mass as in eq. (15) but with lower mean density, allowing the collapse calculation to be spherically symmetric.

### 3.2. Collapse of a dispersing system

Integration of eq. (15) from 0 to r gives the mass within radius r at time t,

$$M(<r,t) = M_*(t) + \left[M(<a) - M_*(t)\right]v(t) \qquad (19)$$

where the protostar point mass $M_*(t) \equiv M(0,t)$ is the mass which has accreted at the center by time t, assuming that the gas mass disperses but the stellar mass does not disperse. Here the accretion disk is ignored, assuming that on average the accretion rate of the core onto the disk is equal to the accretion rate of the disk onto the protostar, and assuming that the mass of the disk is negligibly small compared to that of the core and the protostar.

The equation of motion for the pressure-free collapse is then

$$\ddot{r}(r,t) = -\frac{GM(<r,t)}{r^2} \qquad (20)$$

with M(<r,t) from eq. (19).

In eq. (20), the thermal pressure of the collapsing gas is neglected, on the grounds that the competition between infall and dispersal is most important in setting the stellar mass. This neglect leads to a significant error at the earliest stage of the infall for the nonsingular isothermal sphere, where the small zone of nearly uniform central gas falls in all at once rather than gradually. However for the subsequent isothermal collapse, the time required for each mass shell to collapse to the center is still approximately the free-fall time calculated from the average interior density of the shell (Larson 2003). Numerical calculations of the collapse of an overdense isothermal sphere show very little difference between isothermal and pressure-free mass-accretion rates (Ogino, Tomisaka, & Nakamura 1999).

The rotation of the initial gas is also neglected, since on the size scale considered here, spectral line velocity gradients indicate that rotational energy comprises at most a few percent of the energy budget of a dense core (Goodman et al 1993). Such rotation is important in forming the circumstellar disk (Terebey, Shu & Cassen 1984), but in the present calculation the protostar and disk are considered as a single unit.

The solution to eqs. (19) and (20) is nearly the same as the free fall solution (Hunter 1962, Spitzer 1968), but for this dispersing system more steps are needed to obtain the protostar mass. Defining the parameter $\beta$ by $r \equiv a \cos^2 \beta$, one obtains

$$\beta + \sin\beta \cos\beta = \left(\frac{2GM(<a)}{a^3}\right)^{1/2} \int_0^t dt'\, F(t') \qquad (21),$$

where the dimensionless function of time F is defined by

$$F \equiv \left[\mu + (1-\mu)\nu\right]^{1/2} \qquad (22).$$

and where $\mu$ is the ratio of protostar mass to the initial mass within a:

$$\mu \equiv \frac{M_*}{M(<a)} \qquad (23).$$

Here F and $\mu$ each lie in the range 0 to 1. Eq. (21) reduces to the free fall solution when there is no dispersal, i.e. when $\nu=1$.

3.3. Time evolution of protostar mass

In the presence of dispersal, the "infall time" $t_f(a)$ for a shell of initial radius $a$ to fall to the center exceeds the corresponding free fall time $t_{ff}(a)$ given in eq. (10), since the decreasing mass interior to the shell exerts less gravitational pull on the shell than if there were no dispersal. These two times are related by eq. (21) when $\beta = \pi/2$,

$$t_{ff}(a) = \int_0^{t_f(a)} dt' F(t') \qquad (24).$$

The mass added to the protostar by an infalling shell can be written

$$dM_* = dt_f \frac{dM}{dt_f} \nu = dt_f \frac{dM}{dt_{ff}} \frac{dt_{ff}}{dt_f} \nu \qquad (25)$$

as long as the initial mean density decreases monotonically outward, so that infalling mass shells do not cross (Hunter 1962). Eq. (25) accounts for the dispersal of interior gas through the term $dt_{ff}/dt_f$ and for the dispersal of shell gas through the term $\nu$. Henceforth the rates of mass increase are denoted "free fall rate" $dM/dt_{ff}$, "infall rate" $dM/dt_f$, and "accretion rate" $(dM/dt_f)\nu$.

To evaluate eq. (25) the free fall rate $dM/dt_{ff}$ is obtained from the initial cloud model, as in Section 2.5, and the argument $t_{ff}$ of $dM/dt_{ff}$ is related to the infall time $t_f$ using eq. (24).

Differentiating eq. (24) gives the derivative $dt_{ff}/dt_f = F$. Then integrating eq. (25) over infall times $t_f$ gives the protostar mass as a function of time,

$$M_*(t) = \int_0^t dt_f \frac{dM}{dt_{ff}} F \nu \qquad (26).$$

The protostar mass $M_*(t)$ is thus the mass contributed by all the shells whose infall times $t_f$ range from 0 to t, where each shell contributes the fraction $\nu(t_f)$ of its initial mass.

The mass $M(<a)$ in eqs. (19) and (23) can also be written as $M(t)$, where t is the infall time of the shell initially at a, in a form more convenient for calculation of the protostar mass:

$$M(t) = \int_0^t dt_f \frac{dM}{dt_{ff}} F \qquad (27).$$

Thus in eq. (27) $M(t)$ is the initial mass of all the shells whose infall times $t_f$ range from 0 to t. Note that $M(t)$ gives the initial mass as a function of its infall time, and not the time evolution of the initial mass.

The ratio of protostar to initial mass in eq. (23) can now be written

$$\mu(t) = \frac{M_*(t)}{M(t)} \qquad (28).$$

It is possible to solve eqs. (26), (27), and (28) for $\mu$, M, and $M_*$ once $dM/dt_{ff}$ is given by an initial cloud model and $\nu$ is given by a model of how much gas survives dispersal. An input version of $\mu(t_f)$ is used to compute $M_*(t)$ and $M(t)$ from eqs. (26) and (27), yielding an output version of $\mu(t)$ from the ratio $M_*(t)/M(t)$. The output version is then substituted for the next input version, and the process is repeated until the input and output versions are indistinguishably different.

## 4. RESULTS

### 4.1. Sample calculation

Here the protostar mass and luminosity are calculated as functions of time for specific input parameters, to show how the equations of Section 3 are implemented and to illustrate two characteristic results.

### 4.2. Protostar mass

The protostar mass $M_*(t)$ in eq. (26) and the initial mass whose infall time is t, M(t) in eq. (25), are each written in terms of a mass scale and a dimensionless integral by obtaining $dM/dt_{ff}$ from Section 2.5 and $\nu$ from eq. (18),

$$M_*(t) \equiv M_{*0} m_*(t) \qquad (29)$$

$$M(t) \equiv M_{*0} m(t) \qquad (30)$$

where the protostar mass scale is

$$M_{*0} \equiv \left(\frac{2}{3\pi}\right)^{3/2} \frac{4b\sigma^3 t_d}{G} = 1.31\ \mathrm{M_O} \left(\frac{T}{10\ \mathrm{K}}\right)^{3/2} \left(\frac{t_d}{0.1\ \mathrm{Myr}}\right) \quad (31),$$

the protostar mass integral is

$$m_*(t) \equiv \int_0^{t/t_d} d\theta_f \frac{dm}{d\theta_{ff}} F \exp(-\theta_f) \quad (32),$$

and the initial mass integral is

$$m(t) \equiv \int_0^{t/t_d} d\theta_f \frac{dm}{d\theta_{ff}} F \quad (33).$$

Here the integration variable is the dimensionless infall time $\theta_f \equiv t_f/t_d$. One repeats the integrations of eqs. (32) and (33) until the input and output versions of $\mu = m_*/m$ converge, typically in 4 or 5 iterations. The integration is extended until t = 1 Myr or until t equals the infall time of the radius $a_{max}$ which encloses the initial Jeans mass, whichever comes first.

The outcome of each calculation of $M_*(t)$ depends on the parameters $\sigma$ and $t_d$ in the mass scale $M_{*0}$ in eq. (31), and on two dimensionless ratios--the background-to-peak density contrast f, and on the ratio $t_d/t_0$ of dispersal and peak density free-fall times. These ratios, or others formed from them, delimit the two types of accretion history, as discussed in Section 4.4.

Figure 3 shows $M_*(t)$ and $M(t)$, and Figure 4 shows their ratio $\mu$ for the parameters $T=10$ K, $n_0=10^6$ cm$^{-3}$, $n_u=10^3$ cm$^{-3}$, and $t_d=0.1$ Myr. The zero of time is the start of the infall, but in this pressure-free calculation the start of accretion is delayed by the free-fall time of the peak density, $t_0$. During this short period the gas is collapsing but not yet accreting.

The protostar mass $M_*(t)$ in Figure 3 increases rapidly as the protostar begins its accretion, due to the infall of the nearly uniform portion of the initial density profile. The mass then approaches a time-independent value of 0.47 $M_O$ within a few 0.1 Myr. This well-defined value arises because the fraction of infalling gas which survives dispersal, $\nu$, decreases with time faster than the infall rate, $dM/dt_f$, increases with time. Thus the accretion rate, $\nu\, dM/dt$, decreases with time and the total accreted mass levels off to a nearly constant value. If in addition the background infall time is sufficiently long compared to the dispersal time, a negligible amount of infalling background gas survives dispersal by the time it reaches the protostar.

When the dispersal time $t_d$ is reduced, less gas reaches the protostar and the final protostar mass is also reduced. For example when $t_d=0.068$ Myr with all other parameters the same, the final protostar mass is 0.26 $M_O$, close to the peak of the IMF (Chabrier 2005, Kroupa 2002).

While the protostar mass levels off, the mass whose infall time is t, M(t), continues to increase with time, because the initial mass increases with radius and thus with infall time. Thus the ratio $\mu$ in Figure 4 decreases with time. Figure 4 also shows both the input and output versions of $\mu$ in the last iteration used, indicating a well-defined solution.

## 4.3. Accretion luminosity

The accretion luminosity of the protostar is given by

$$L_{acc} = \frac{GM_* \dot{M}_*}{R_*} \qquad (33)$$

in spherical symmetry (Adams & Shu 1985), with the protostar radius $R_*$ taken to be 3 $R_O$ (Stahler 1988). Then eqs. (29) and (31) yield

$$L_{acc} \equiv L_{acc0} \ell_{acc} \qquad (34)$$

where the accretion luminosity scale is

$$L_{acc0} = (4b)^2 \left(\frac{2\sigma^2}{3\pi}\right)^3 \frac{t_d}{GR_*} \qquad (35)$$

or

$$L_{acc0} = 8.1 \, L_O \left(\frac{T}{10 \text{ K}}\right)^3 \left(\frac{t_d}{0.1 \text{ Myr}}\right) \left(\frac{R_*}{3R_O}\right)^{-1} \qquad (36)$$

and the dimensionless accretion luminosity is

$$\ell_{acc} \equiv m_* \frac{dm}{d\theta_{ff}} F\nu \qquad (37).$$

The time history of the accretion luminosity is shown in Figure 5 for the same parameters used in the mass calculation shown in Figures 3 and 4. The accretion luminosity has a short "spike" and a long "tail," due to the rapid rise and rolloff of the protostar mass history as in Figure 3. If there were no dispersal, the accretion luminosity would increase monotonically and would have no spike. This FWHM duration of the spike, 0.11 Myr, is similar to estimates of the duration of the Class 0 phase, 0.03 Myr (André, Ward-Thompson & Barsony 2000) to 0.16 Myr (Evans et al 2008).

The typical accretion luminosity of this model is similar to the typical bolometric luminosity of accreting young stellar objects. The young stellar objects in Perseus, Serpens, and Ophiuchus observed by the Spitzer "c2d" survey have a 90% completeness limit on bolometric luminosity of 0.05 $L_O$ (Evans et al 2008, Harvey et al 2007), and the bolometric luminosity distribution of the 128 Class 0 and Class I objects detected above this limit has median 1.0 $L_O$ (Evans et al 2008). In Figure 5, the distribution of accretion luminosity above this same limit has a comparable median of 0.8 $L_O$.

As dispersal time is reduced, the accretion luminosity more rapidly reaches a lower peak of shorter duration. In the above example of $t_d$=0.068 Myr which gives a final protostar mass matching the peak of the IMF, the peak luminosity is 5.1 $L_O$ and its FWHM duration is 0.07 Myr, again similar to the duration of the "Class 0" phase.

The standard model of isolated star formation (Shu, Adams & Lizano 1987) and other models of steady accretion predict protostar luminosities greater than are typically observed (McKee & Ostriker 2007, Kenyon et al 1990). A suggested solution to this "luminosity problem" is that disk instabilities cause short episodes of high luminosity separated by long periods of low luminosity (Vorobyov & Basu 2005; see also Tassis & Mouschovias 2005). The time history of Figure 5 can be viewed as a single cycle of such high and low luminosity. Thus it is possible that the broad range of observed luminosities is due not only to nonsteady accretion from the disk to the star, but also to nonsteady accretion from the core to the star-disk system, as in the present model.

4.4. Two types of accretion

The protostar mass described in Sections 4.2 and 4.3 can increase in two characteristic ways, depending on the dispersal time $t_d$ in relation to the free-fall times of the IS+U gas.

The dispersal time scale $t_d$ sets the level of competition between the rates of infall and dispersal. If $t_d$ is much shorter than the free fall time $t_0$ of the innermost mass shell, the gas will disperse long before it can accrete. If $t_d$ is much longer than the free fall time $t_u$ of the outermost shell, the accretion "runs away" until nearly all of the initial Jeans mass has accreted onto the protostar. When $t_d$ lies between these times a substantial but limited amount of dense gas can accrete. In this case the innermost shells lose little gas because their infall times are shorter than $t_d$, while the outermost shells are mostly dispersed because their infall times are longer than $t_d$.

Figure 6 shows accretion histories illustrating these points, with standard parameter values, and dispersal times $t_d$ = 0.034, 0.17, 0.34, and 0.55 Myr. The free-fall times of the innermost and outermost gas are $t_0$ = 0.034 Myr and $t_u$ = 1.1 Myr, respectively, so the dispersal times in Figure 6 span a large fraction of the range of free-fall times.

The mass accretion histories in Figure 6 can be distinguished by their behavior at $t=t_u=1$ Myr. There, the slope $dM_*/dt$ is increasing with time for $t_d$ = 0.55 Myr, constant for $t_d$ = 0.34 Myr, and decreasing for $t_d$ = 0.17 and 0.034 Myr.

The case of increasing slope is denoted "runaway" accretion in analogy with the rapid growth of planetesimals from a background of small bodies (Goldreich, Lithwick & Sari 2004). This type of accretion arises because the relatively long dispersal time allows first core gas and then background gas to fall in. As background gas begins to fall in, the infall rate $dM/dt_f$ increases rapidly with time because the background is more nearly uniform. As a result the accretion rate $(dM/dt_f)\nu$ begins to increase with time and the accretion "runs away." If such accretion continues until the entire Jeans mass is either accreted or dispersed, the resulting protostar mass in Figure 6 is 5.4 $M_O$ for $t_d = 0.55$ Myr. For such accretion the luminosity profile does not have the spike of Figure 5 but instead increases monotonically with time. Such high accretion rate is suggestive of conditions for massive star formation, as discused in Section 5.

Among the cases of decreasing slope, those with shallowest slope are of interest because then the protostar mass at $t = t_u$ is nearly constant. These cases are denoted "self-limiting" accretion when the slope at time $t_u$ is less than or equal to 0.05 $M_*(t_u)/t_u$. The coefficient 0.05 is chosen arbitrarily to give a negligibly small rate of increase. For such self-limiting accretion the protostar mass arises largely from within the dense core, and is independent of the mass and density profile of the background gas.

4.5. Dependence of protostar mass on density and dispersal time

A sequence of accretion histories was calculated to define the range of parameters which gives protostar masses which are self-limiting and which lie within the span of observed brown dwarfs and stars. The peak densities $n_0$ range from $10^5$ to $10^6$ cm$^{-3}$, and the background densities $n_u$ range from $10^3$ to $10^4$ cm$^{-3}$, matching properties of nearby star-forming clouds (Enoch et al 2007). The dispersal time $t_d$ was varied from 0.02 to 0.55 Myr to span a significant fraction of the range of peak and background free-fall times. The temperature was assumed to be 10 K. The resulting conditions on dispersal time can be approximated as

$$t_d \geq t_0 \tag{38}$$

to achieve protostars with mass greater than or equal to 0.05 $M_O$, and

$$t_d \leq 0.2 t_u \tag{39}$$

to ensure that the final protostar mass is well-defined, i.e. the accretion is self-limiting as described in Section 4.4 above.

The range of free fall times in eqs. (38) and (39) implies that the ratio of densities $n_0/n_u$ must exceed $(0.2)^{-2} = 25$. Thus a sufficiently high peak-to-background density contrast is necessary to obtain well-defined protostar masses from the competition between dispersal and infall.

The variation of final protostar mass with dispersal time is shown in Figure 7 for peak and background densities $10^6$ and $10^3$ cm$^{-3}$. This mass increases in roughly linear fashion with $t_d$ for both self-limiting and runaway accretion, and lies close to the curve of initial gas mass whose free-fall time equals $t_d$, based on eqs. (6) and (12). Thus the final protostar mass can be understood approximately as the initial IS+U mass whose free-fall time equals $t_d$, or from eq. (14),

$$M_* \approx \frac{8\alpha_*^{3/2} \sigma^3 t_d}{\pi G} \qquad (40).$$

where $\alpha_*$ is the value of $\alpha$ at the radius whose free-fall time equals $t_d$. As noted in Section 2.1, $\alpha \approx 1.3$ for the linear portion of the dependence of mass on free-fall time. In Figure 7 this portion extends roughly from log $(t_d/Myr)$=-1.0 to -0.5.

In Figure 7, the Jeans mass exceeds the runaway masses, which in turn exceed the self-limiting masses, as expected. However, the self-limiting masses can be greater than or less than the core mass, depending on the value of $t_d$. The protostar mass does not have a fixed relation to the core mass, unless $t_d$ lies in a relatively narrow range.

4.6. Matching protostar and core properties

Figure 7 shows that the IS+U model with competing infall and outflow can produce a wide range of protostar masses, depending on the dispersal time scale. In this section star and core properties are discussed which limit the typical dispersal time to a relatively narrow range 0.05-0.1 Myr, or equivalently which limit the ratio $t_d/t_{core}$ to the range 0.4-0.8. With this limitation, the IS+U model reproduces four properties of protostars and cores.

*Typical protostar mass.* If cores form protostars in roughly one-to-one fashion, the most common core properties should produce the most common protostar mass, about 0.3 $M_O$ (Chabrier 2005, Kroupa 2003). For the most common core temperature, ~10 K (JMA), and for the peak and background densities $10^6$ and $10^3$ cm$^{-3}$, adopted earlier, the protostar mass 0.26 $M_O$ lies near the peak of the IMF and arises from dispersal time 0.068 Myr according to Section 3.5.

*Ratio of protostar and core mass.* The typical ratio ε of protostar and core mass should be of order one-half, as suggested by recent studies of core mass distributions (Stanke et al 2006, Alves, Lombardi & Lada 2007, Enoch et al 2008). For standard parameters and the core definition in Section 2.2, the core mass is 0.68 $M_O$ according to Section 2.3, so the model ratio of protostar to core mass is ε = 0.4, near the expected value.

*Correlation of protostar and core masses.* Protostar masses should correlate with core masses, due to the resemblance of the core mass distribution to the IMF in nearby star-forming regions. The protostar and core mass each scale with velocity dispersion as $\sigma^3$, as in eqs. (14) and (40). Thus their masses correlate if the range of σ is sufficiently large while the ranges of other parameters are sufficiently small. From eqs. (14) and (40), these other parameters are the ratio of dispersal and core free-fall times $t_d/t_{core}$, and the ratio of the dimensionless mass per radius α for the initial radii which enclose the protostar and core mass:

$$\frac{M_*}{M_{core}} \approx \left(\frac{\alpha_*}{\alpha_{core}}\right)^{3/2} \frac{t_d}{t_{core}} \qquad (41).$$

The most influential of these parameters is the ratio of core and dispersal times, since $\alpha_*$ and $\alpha_{core}$ are nearly equal when the core and dispersal times are comparable. For standard values of $n_0$ and $n_u$, Figure 7 indicates that $M_{star}$ is confined to the range 0.2-0.6 $M_{core}$ if $t_d$ lies in the range 0.05-0.1 Myr, or equivalently if the ratio $t_d/t_{core}$ lies in the range 0.4-0.8.

*Range of protostar mass.* The range of protostar masses should be substantial, if the model is to account for the masses of most stars. The minimum velocity dispersion comes from the coldest cores, which have central kinetic temperature based on $NH_3$ line observations as low as ~7 K (Crapsi et al 2007) and negligible nonthermal motions. In clouds with a broad range of stellar masses, such as L1641 in Orion, the highest dense gas temperatures approach 100 K. Of the 339 mapped $NH_3$ cores in the JMA database, nine have well-determined temperatures exceeding 50 K. For the temperature range 7-50 K, the corresponding mass range is a factor of $(50/7)^{3/2} =19$. For this comparison, the standard densities of Section 2.1 are assumed to be fixed, and nonthermal motions are neglected. Relaxation of these assumptions is discussed in Section 5.5.

In summary, the foregoing requirements on protostar and core masses can be satisfied by gas temperatures T=7-50 K, dispersal times 0.05-0.1 Myr, and peak and background densities $10^6$ and $10^3$ $cm^{-3}$, as shown in Figure 8. Then gas with typical temperature 10 K and dispersal time 0.07 Myr produces a protostar whose well-defined mass matches the peak of the IMF. The protostar masses span the range 0.08 - 5 $M_O$, protostar mass correlates with core mass, and their mass ratio is typically 0.4 ± 0.2.

4.7. Detectable life of a core with a protostar

The dispersal time scale of 0.07 Myr, obtained above in order to reproduce a typical protostar mass from typical initial density and temperature, can now be used to estimate the detectable lifetime of a core which forms such a protostar. The time for the density to decrease from its peak to its background value, is from eq. (18),

$$t_n = t_d \ln f^{-1} \qquad (42)$$

or 0.48 Myr for the standard value $f=10^{-3}$. Similarly the time for the column density to decrease from its peak to background value is

$$t_N = t_d \ln f^{-1/2} \qquad (43)$$

or 0.24 Myr, since the IS+U column density is proportional to the square root of the volume density.

Detection of cores by their dust emission or absorption depends primarily on their column density, while detection of cores by their molecular line emission depends on both column density and volume density. Taking the column density disappearance time $t_N$ as more realistic indicates that the detectable lifetime of a core with a protostar is typically 3-4 dispersal time scales. The lifetime derived from the observations cited in Section 1.2 is typically 0.2-0.3 Myr, and thus is in good agreement with these estimates.

On the other hand, it is less clear from observations whether the local environment of a core disperses as quickly as the core itself, as in the dispersal model adopted here. A "Class I" protostar is seen about half the time without an accompanying core (Jørgensen et al 2008). But such a protostar is typically projected on several magnitudes of visual extinction, extended over at least ~0.1 pc. Such "extinction cores" (KJD) have density similar to the IS+U background density, ~ $10^3$ cm$^{-3}$, but have greater column density than a single IS+U unit, typically ~2 magnitudes.

This column density difference could be understood if the typical observed cloud contains many Jeans masses, and thus contains at least a few Jeans masses along the line of sight.

Then dispersal of a single IS+U unit would remove the core but would still leave a substantial projected background, due to other cloud gas along the line of sight.

## 5. DISCUSSION

### 5.1. Relation to other models

The model presented here resembles early models of star formation (Larson 1969, Shu 1977) in its calculation of gravitational collapse and accretion from a spherically symmetric distribution of initial density, but differs because it includes dispersal of the initial gas as a competing process. In a pure infall model the final protostar mass is simply the initially available gas mass. But the competition between infall and dispersal selects a protostar mass from a continuum of possible values, depending on the relative values of infall and dispersal time.

This model also differs from most models of initial conditions for infall, because the initial density profile has both a steep inner slope and a shallow outer slope, to better represent observed dense cores and their surrounding gas. This structure leads to slow accretion of dense core gas followed by faster accretion of surrounding gas, until dispersal shuts off the process. Then the dispersal time selects not only the final protostar mass, but also the accretion rate characteristic of low-mass or massive star formation.

### 5.2. Dispersal time and core free-fall time

Matching model predictions with observed correlations between core and protostar masses requires that $t_d$ lie in a narrow range with respect to the core free-fall time, $t_d$=(0.4-0.8)$t_{core}$, as described in Section 4.6. This similarity of time scales suggests that the physical basis of the dispersal is closely related to the gravitational unbinding of the core. The "escape time" $t_{esc}$ can be defined as the time for gas at a given radius to travel its own radius at its escape speed. This time is nearly the same as the free-fall time from the same radius, since

$$t_{esc} = r\left(\frac{r}{2GM(<r)}\right)^{1/2} = \left(\frac{2}{\pi}\right)^{1/2} t_{ff} \qquad (44).$$

Thus the similarity of $t_d$ and $t_{core}$ deduced earlier implies that $t_d$ must also be similar to the escape time from the core radius. Specifically, $t_d$=(0.5-1.0)$t_{esc}$.

For standard parameter values, the escape speed from the initial core radius is 0.4 km s$^{-1}$. This speed is much less than the typical maximum outflow speed of 11 km s$^{-1}$ (Wu, Huang & He 1996). This comparison suggests that much of the core gas disperses more slowly than indicated by outflow speeds. If so, one may speculate as in Section 3.1 that in the typical case of self-limiting accretion, the outflow rapidly disperses enough core gas to unbind the remainder, which then disperses more slowly, at a speed closer to the escape speed.

5.3. Magnetic fields and turbulent motions

The magnetic fields evident in Zeeman splitting (Crutcher 2007) and submillimeter polarization (Hildebrand et al 2000) and the turbulent motions seen in supersonic line widths (Larson 1981) are believed to have a strong influence on the large-scale structure of molecular clouds seen in both observations and simulations (Larson 2003). Similarly, flows associated with ambipolar diffusion, and with HD and MHD turbulence are likely to be important in forming and condensing cores and their immediate surroundings. Significant turbulent motions are also common in cores associated with the most massive stars (Beuther et al 2007).

However, for starless cores in well-studied nearby regions, thermal physics appears to dominate over magnetic and turbulent effects (Larson 2005). The nonthermal velocity dispersion derived from the widths of $NH_3$ or $N_2H^+$ lines is typically less than the thermal velocity dispersion in isolated cores in nearby dark clouds (Myers 1983, Fuller & Myers 1992) and in nearby young clusters such as L1688 in Ophiuchus (Ward-Thompson et al 2007, Fig. 6). This property may be expected from large-scale simulations in which turbulent motions decay within a dynamical time scale (MacLow & Klessen 2004).

If some of these observed nonthermal motions are due to magnetic effects, the corresponding magnetic energy density must be less than the thermal energy density. The magnetic field can strongly limit the rate of core condensation by ambipolar diffusion in two dimensions, but once the core is a distinct entity, its further evolution and collapse are very similar to those in the competition between thermal pressure and gravity. In the standard case of magnetized sheet collapse considered by Adams & Shu (2007), the mass accretion rate near the origin has form identical to that for the collapse of the unmagnetized SIS, but with a coefficient greater by a factor 1.8.

Consequently, the models in this paper are based on thermal physics and self-gravity, with turbulent and magnetic effects ignored. Inclusion of turbulent and magnetic pressure at the level indicated by observations are expected to increase the core velocity dispersion and mass, as discussed by Larson (2005), but not to change the basic qualitative features of the results.

5.4. Runaway accretion

The runaway accretion described in Section 4 is of interest for models of massive star formation (e.g. Zinnecker & Yorke 2007), since the final protostar mass is not limited by dispersal, and can reach massive star values for standard parameters. This subsection discusses background geometries and dispersal models which enable runaway accretion.

Runaway accretion occurs when the accretion rate, $(dM/dt_f)\nu$ in eq. (25), increases with time. For the IS+U configuration with exponentially decreasing survival, a sufficiently high contrast between core and background density ensures that core infall occurs much sooner than background infall. If in addition the dispersal time is comparable to the core infall time, the infall rate $dM/dt_f$ is approximately constant during the core infall. Then the nearly constant infall rate multiplied by the exponentially decreasing survival function gives a declining accretion rate and thus self-limiting accretion. However if the dispersal time is too long compared to the background free-fall time, then a significant amount of infalling background gas survives dispersal. As this nearly uniform gas approaches the protostar, the infall rate $dM/dt_f$ increases rapidly, because uniform gas has the steepest possible infall rate. This infall adds mass to the protostar faster than it is dispersed, and so the accretion runs away.

For exponentially decreasing survival as in eq. (18), background density profiles which fall off as negative powers of radius cannot run away. For example if the initial density profile beyond the core follows $n \sim a^{-2}$ as for the isothermal sphere with f=0 in eq. (1), no runaway is possible because the infall rate $dM/dt_f$ is nearly constant with time while the survival function $\nu$ decreases with time. For background density declining as in a logotropic sphere (McLaughlin & Pudritz 1996), with $n \sim a^{-1}$, the free fall rate $dM/dt_{ff}$ increases with free fall time approximately as $t_{ff}^3$. This dependence is much steeper than for the SIS, but still not steep enough to run away against exponentially decreasing survival.

Similarly, backgrounds which are uniformly extended in only one dimension, as in a uniform or isothermal cylinder (Ostriker 1964), or in two dimensions, as in a uniform or isothermal layer (Spitzer 1942), can run away only until gas along their short dimensions is used up. After that time, each such free fall rate will be less than the corresponding free fall rate of its spherically-averaged mass distribution. This spherically-averaged free fall rate depends on free-fall time as $t_{ff}^0$ for the cylinder and as $t_{ff}^3$ for the layer, essentially the same as for the SIS and the logotropic sphere, respectively. Thus core backgrounds having the geometry of layers or filaments cannot support runaway accretion if the survival function declines with time exponentially.

These considerations suggest that significant runaway accretion can occur only for sufficiently long dispersal time, and for initial backgrounds having nearly uniform structure, with significant extent in all three dimensions. This combination of requirements may be difficult to fulfill in most star-forming regions, suggesting that the massive star formation expected from runaway accretion is confined to unusual initial conditions. These conditions may be found only in the densest parts of cluster-forming regions and not in their more extended filaments.

On the other hand, the exponentially decreasing survival function in eq. (18) may be unrealistically steep. It is based on observations of outflows with poorly known ages and not on a detailed physical model. If a more realistic description of core dispersal becomes available, the range of backgrounds which enable runaway accretion may increase. For example, if the outflow solid angle $\Omega_{out}$ increased with time t as $\Omega_{out} = 4\pi/[1+(t/t_d)^{-2}]$, the survival function would be $\nu = 1/[1+(t/t_d)^2]$, allowing runaway accretion for logotropic as well as uniform backgrounds.

However there is no possible runaway for background gas with the density structure of a SIS or of a uniform or isothermal cylinder, since these backgrounds have infall rates which do not increase with time, as noted above. Therefore this model predicts that the mass function of stars in the most filamentary parts of embedded clusters should lack the high-mass stars which require runaway accretion for their formation.

5.5. Core mass and velocity dispersion

The kinetic temperature range 7-50 K applies to cloud complexes with massive stars, such as L1641 in Orion (JMA), and for this range the IS+U model predicts a substantial range of core and protostar masses, with the ratio of maximum to minimum mass ~60, as in Figure 8. However in regions of lower-mass star formation where the core mass function (CMF) resembles the IMF, the range of kinetic temperature is too small to account for the observed range of core masses.

In the Perseus complex Enoch et al (2008) find good agreement between the power-law slopes of the IMF and the CMF. In this study, 70 cores mapped at 1.1 mm by their dust continuum emission also have kinetic temperature measurements from a $NH_3$ line survey (Rosolowsky et al 2007). These core masses span 0.3-4.6 $M_O$, while the core masses predicted by the IS+U model using observed kinetic temperatures, neglecting nonthermal motions, and assuming standard densities, span 0.6-1.2 $M_O$. This mass range is ~7 times smaller than observed. A similar discrepancy is seen in the Pipe nebula (Alves, Lombardi & Lada 2007, Lada et al 2008).

It is possible to match the observed range of Perseus core masses having known temperatures using eq. (7), but only if the nonthermal motions in the $NH_3$ line widths are included in the velocity dispersion, and if the core minimum density is allowed to vary from core to core rather than being fixed at $3 \times 10^4$ cm$^{-3}$. Then the typical velocity dispersion (the quadrature sum of the thermal and nonthermal velocity dispersions) exceeds the purely thermal value by a factor 1.3, and the required range of core minimum density is $1.3 \times 10^4$ cm$^{-3}$ to $1.7 \times 10^5$ cm$^{-3}$.

Thus it is possible for core masses and self-limiting protostar masses to correlate as in Figure 8 for a small range of dispersal times, for fixed core densities, and for a large range of purely thermal velocity dispersion, as in Orion. However, for nearby star-forming regions such as Perseus it is necessary to also take nonthermal motions and varying core densities into account.

5.6. Core masses and the IMF

In the models of this paper, self-limiting protostar masses arise from their parent dense cores with substantial efficiency, provided the outflow dispersal time is comparable to the dense core free-fall time. For regions of low-mass star formation the typical dispersal time required by the model agrees well with the dispersal time indicated by estimates of core lifetimes. With this context, the origin of the IMF becomes a question of the origin of dense core masses, as has been noted by many authors (e.g. Hennebelle & Chabrier 2008).

Some guidance to this problem may come from the structure of young clusters. To achieve the high space density of low-mass protostars in clusters with the IS+U model requires high background densities $n_u$, which ensures short Jeans lengths and also low protostar masses. For peak density and temperature $n_0 = 10^6$ cm$^{-3}$ and 10 K as assumed earlier, and for a hundredfold increase in background density, to $n_u=10^5$ cm$^{-3}$, the Jeans length is reduced to 0.01 pc and the core mass is reduced to 0.2 $M_O$. These properties are appropriate to closely-spaced low-mass star formation as in embedded clusters. At the same time, such high background densities also promote the incidence of runaway accretion because the density contrast between core and background is much less than assumed earlier.

Thus, the two modes of star formation in this model allow a region of fixed temperature and density to make stars of both low and high mass, unlike the simple Jeans instability model. A few instances of runaway accretion may be sufficient to form stars massive enough to provide heating but not massive enough to disperse the entire cluster-forming gas. Then star formation in big clusters like L1641 might differ from that in smaller clusters like NGC1333 or from isolated regions like Taurus, because L1641 has a greater mass of dense background gas.

## 6. SUMMARY

In this paper, an isothermal sphere in a uniform background collapses and disperses to form a protostar. This initial model represents the density structure of observed cores, and also the low-density gas on which they are projected. The background gas available for accretion extends for a Jeans radius, typically a few 0.1 pc. The dense core outer boundary encloses a mass comparable to the peak of the IMF, but this boundary is not a barrier to mass accretion as in the BE sphere.

Accretion is limited by dispersal of surrounding gas. Gas is dispersed primarily by outflow from the forming protostar, and the outflow solid angle increases with time, as shown by Arce & Sargent (2006). Gas may also be dispersed by erosion due to outflows and ionization from young stars outside the core, and by gravitational unbinding once a substantial fraction of the core gas is removed. The gas dispersal is quantified by a time scale $t_d$, which is constrained by observed properties of protostars and cores to have typical value $(0.4\text{-}0.8)t_{core}$, where $t_{core}$ is the free-fall time of the core, typically 0.1 Myr. As a result the protostar mass is set by the competition between infall and dispersal.

Infall with dispersal is calculated in spherical symmetry, assuming negligible gas pressure, rotation, and magnetic fields. The gas which has survived dispersal during its infall does not disperse after it joins the protostar. The calculation predicts the mass and accretion luminosity of the protostar as a function of time.

The protostar accretion has two distinct modes. When the dispersal time is sufficiently small compared to the background infall time $t_u$, $t_d < \sim 0.2 t_u$, the accretion rate decreases with time. Then the protostar mass approaches a constant value, and the accretion is "self-limiting." In this case the protostar mass is approximately the initial mass whose free-fall time is $t_d$. This mass comes mostly from within the core, and does not depend much on the gas beyond the core.

When the dispersal time approaches the background infall time, the net accretion rate increases with time, and the protostar mass comes from both core and background. Such "runaway" accretion resembles that described by some numerical simulations, and may be important for formation of massive stars. Such accretion depends strongly on the structure of the background gas. Runaway accretion is not possible for cores in highly filamentary backgrounds, and the mass distributions of young stars in such backgrounds should therefore lack the very massive stars seen in embedded clusters.

For self-limiting accretion, the mass and accretion luminosity increase rapidly, in a time comparable to $t_d$. The mass levels off to a constant value approximated by eq. (40), while the accretion luminosity has a sharp peak followed by a long, low-luminosity "tail". For conditions which give a protostar mass near the peak of the IMF, the accretion luminosity exceeds 1 $L_O$ for 0.1 Myr, similar to estimates of the duration of the "Class 0" phase.

The dispersal time must lie in the narrow range ~(0.4-0.8)$t_{core}$ in order to have a typical core temperature of 10 K produce the mass 0.3 $M_O$ at the peak of the IMF, and in order to maintain a correlation between self-limiting protostar and core masses, so that their ratio is 0.4 ± 0.2. This small range of $t_d/t_{core}$ suggests that some gas has dispersal speed slower than typical outflow speeds, perhaps due to gravitational unbinding.

Gas with temperature 7-50 K, as seen in $NH_3$ line observations in Orion, yields self-limiting protostar masses ranging from those of brown dwarfs to those of massive stars, 0.08 - 4.9 $M_O$. In nearby clouds with smaller temperature range, it is necessary to allow additional variations in core density and nonthermal velocity dispersion in order to obtain the same correlation between core and protostar masses.

ACKNOWLEDGEMENTS

Helpful discussions with Fred Adams, Hector Arce, Paola Caselli, James Di Francesco, Neal Evans, Eric Feigelson, Gary Fuller, Alyssa Goodman, Tom Hartquist, Doug Johnstone, Jes Jørgensen, Charlie Lada, Chris McKee, Ralph Pudritz, Irwin Shapiro, Frank Shu, and Sue Terebey are gratefully acknowledged. Some material in this paper was developed during the workshop "Star Formation Through Cosmic Time" at the Kavli Institute for Theoretical Physics at the University of California, Santa Barbara.

# APPENDIX. INITIAL MASS AND DENSE CORE RADIUS

For the IS+U initial density model, the mass within a given radius is expressed in terms of the dimensionless function $\alpha(\xi)$ in eq. (6). Integration of eq. (1) gives this function as

$$\alpha \equiv \left[1 + \frac{1}{2\xi}\left(Dd\,\mathrm{Arc\,tan}\frac{\xi}{d} - Cc\,\mathrm{Arc\,tan}\frac{\xi}{c}\right)\right](1-f) + \frac{f\xi^2}{6} \quad \text{(A1)}.$$

If the density ratio f is sufficiently small, $\alpha(\xi)$ is nearly constant for a significant range of $\xi$, as discussed in Section 2.3. In this case the mass within a given radius increases nearly linearly with radius, as in the limiting case of the SIS.

The dense core definition $\rho > \rho_m \equiv (\rho_0 \rho_u)^{1/2}$ combined with eqs. (1) and (2) gives an equation quadratic in $\xi^2$, whose solution is eq. (8), in terms of the dimensionless function $\delta$:

$$\delta \equiv \left\{\left(\frac{1-f}{1-f^{1/2}}\right)\left[\frac{1-11g+\left(1+98g+g^2\right)^{1/2}}{2}\right]\right\}^{1/2} \quad \text{(A2)}$$

with

$$g \equiv \frac{f^{-1/2}-1}{f^{-1}-1} \quad \text{(A3)}.$$

Here $\delta$ approaches unity as f approaches zero. For the standard value $f=10^{-3}$, $\delta=1.17$.

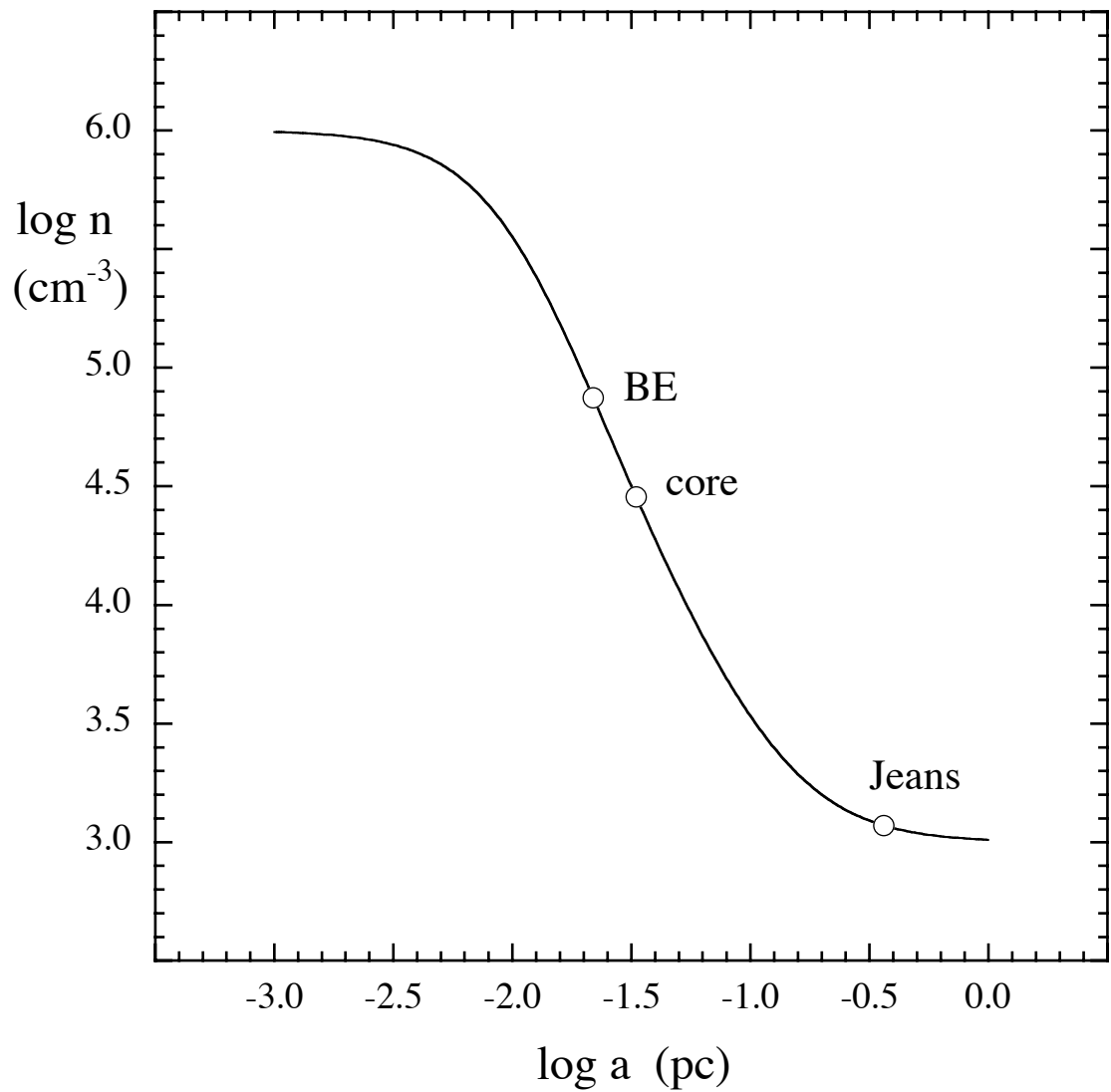

Figure 1. Density n as a function of radius a for an isothermal sphere in a uniform medium, with temperature $T = 10$ K, peak density $n_0 = 10^6$ cm$^{-3}$, and background density $n_u = 10^3$ cm$^{-3}$. Circles indicate the radius of the critically stable Bonnor-Ebert sphere, the dense core with minimum density $(n_0 n_u)^{1/2}$, and the Jeans mass at the background density.

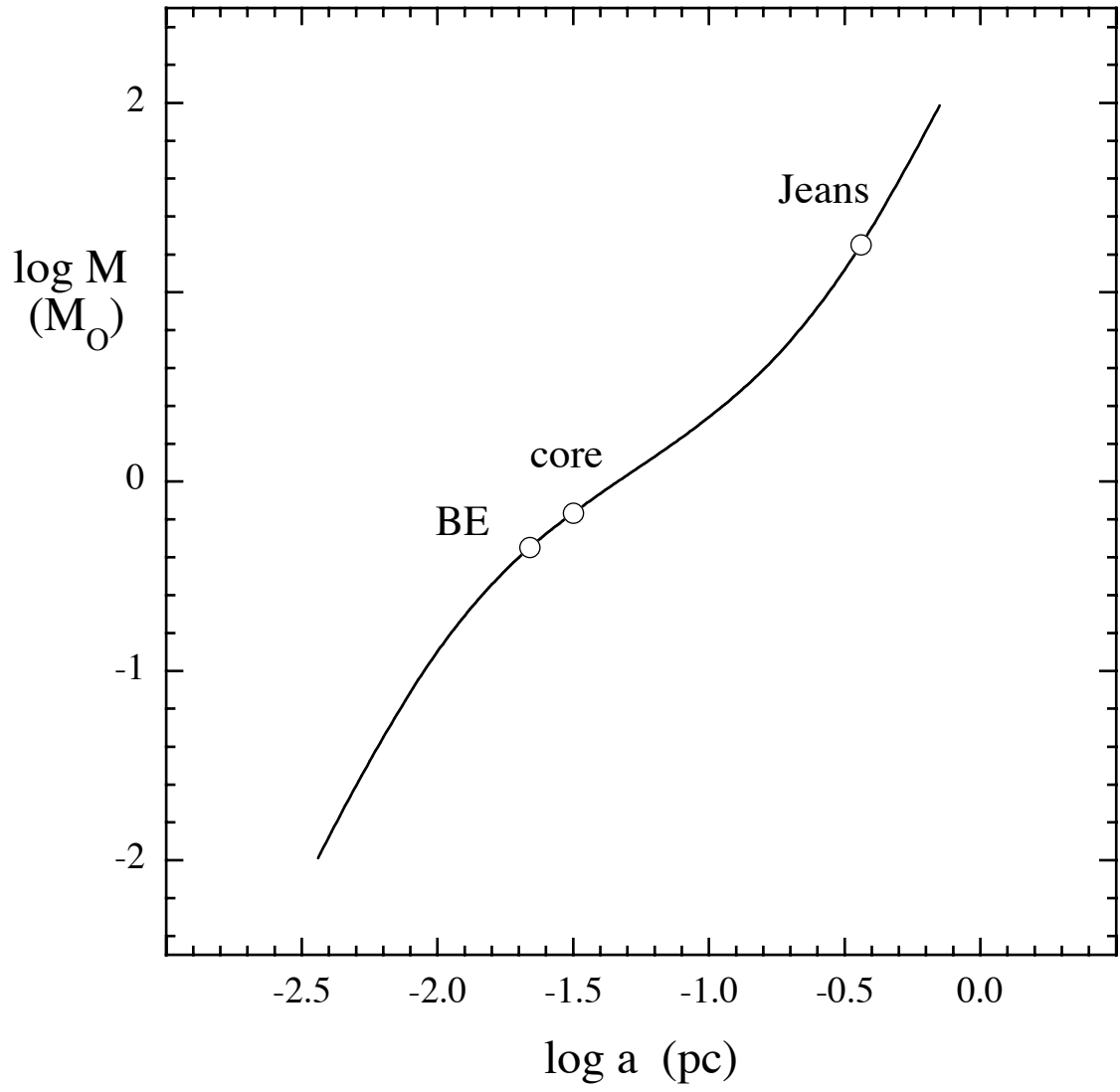

Figure 2. The mass M enclosed by radius a for an isothermal sphere in a uniform medium, with same parameters and labels as in Figure 1.

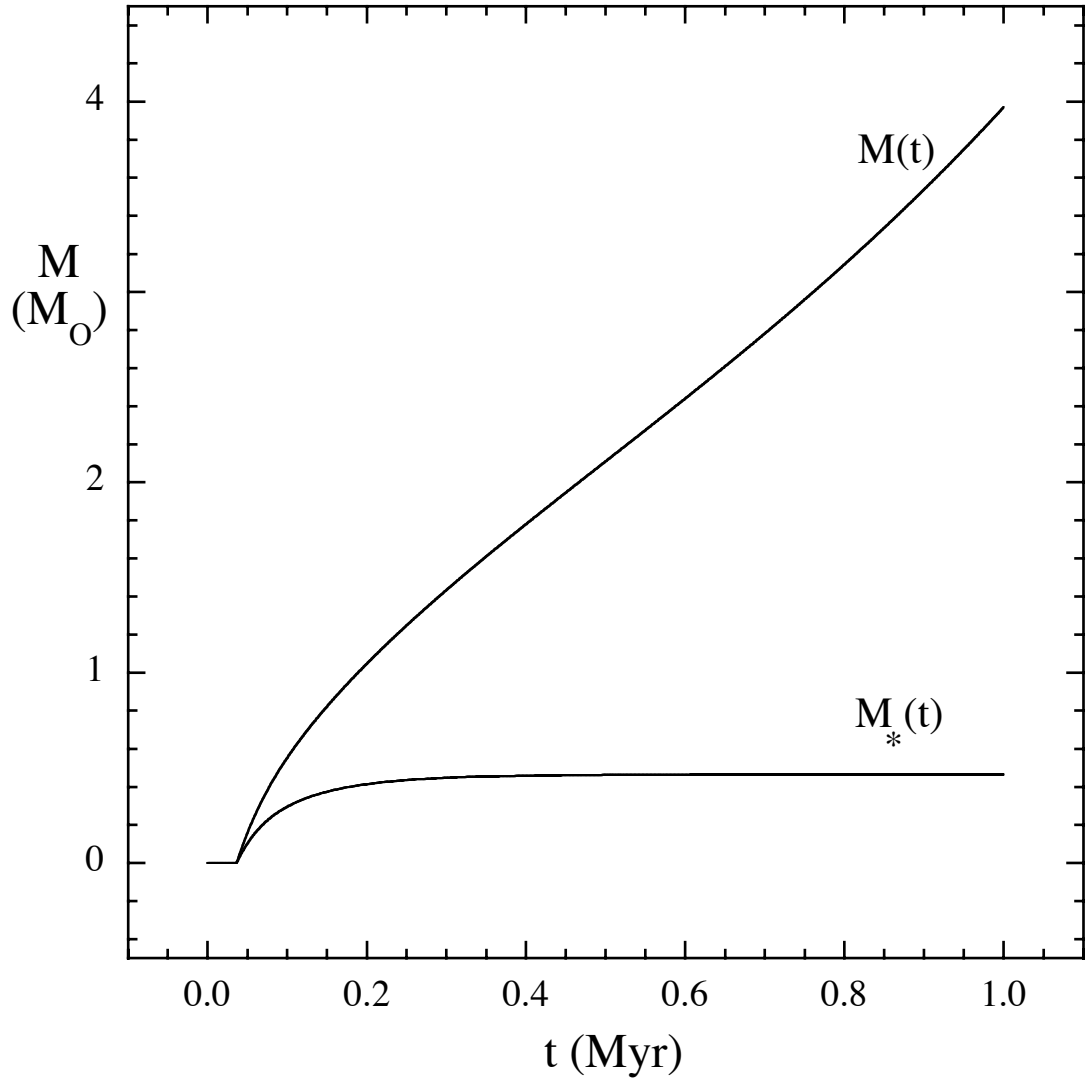

Figure 3. Protostar mass $M_*(t)$ as a function of time t since the start of infall, and initial mass $M(t)$ whose infall time is t, for initial conditions T=10 K, $n_0=10^6$ cm$^{-3}$, $n_u=10^3$ cm$^{-3}$, and $t_d=0.1$ Myr. The protostar mass increases rapidly and then levels off to a time-independent value.

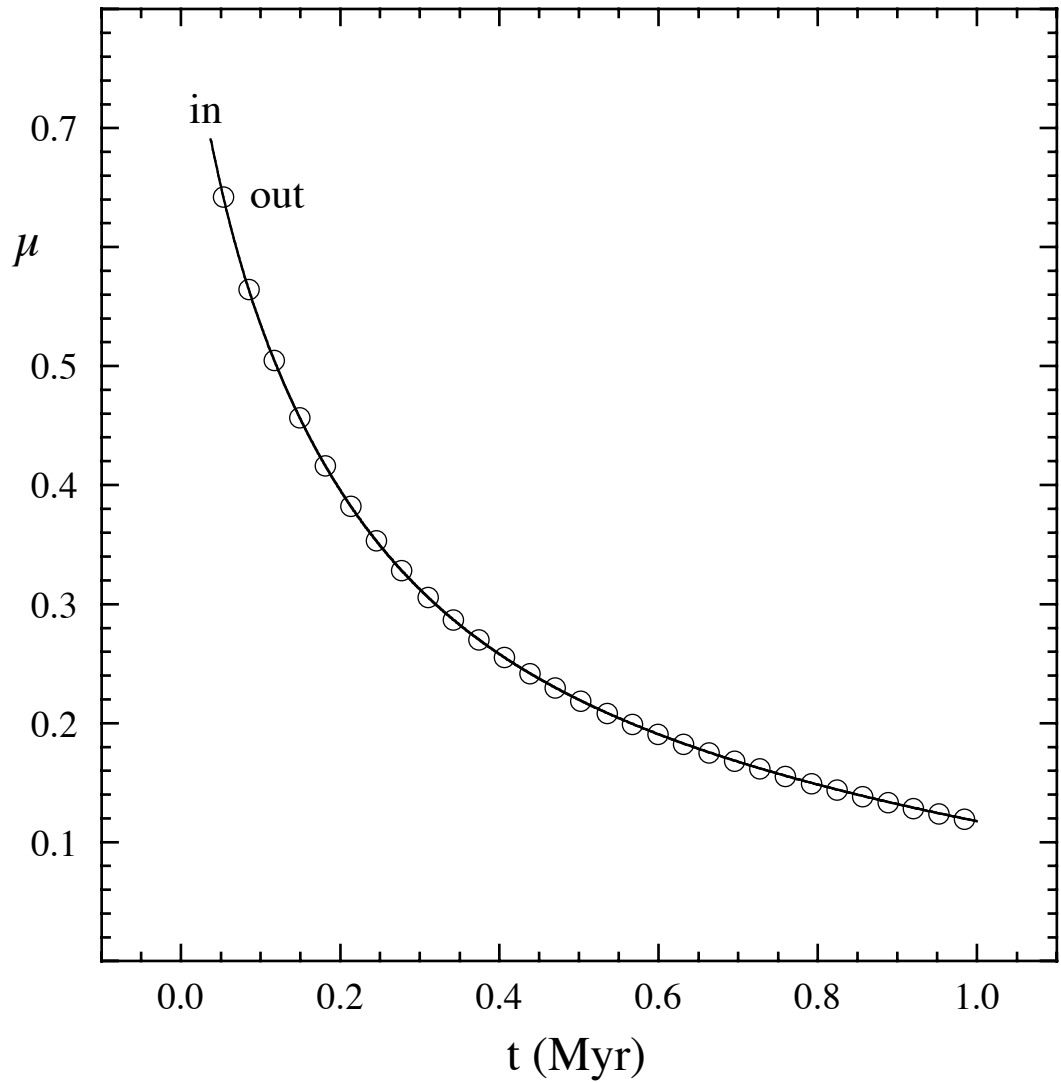

Figure 4. Mass ratios $\mu = M_*(t)/M(t)$ as a function of time t, showing good agreement between input (solid line) and output values (circles) for the calculations in Figure 3.

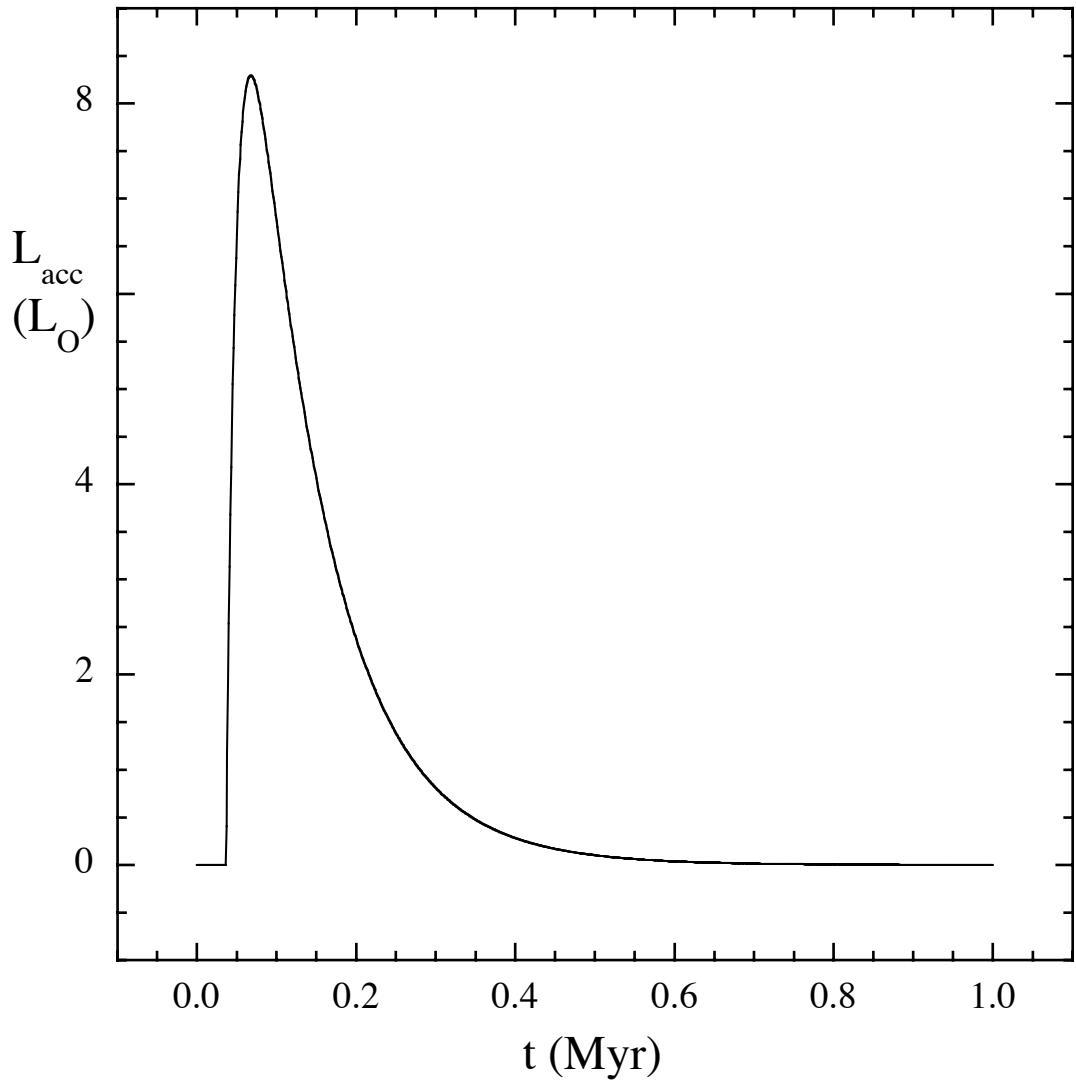

Figure 5. Accretion luminosity of a protostar, for parameters T=10 K, $n_0=10^6$ cm$^{-3}$, $n_u=10^3$ cm$^{-3}$, and $t_d=0.1$ Myr as in Figures 3 and 4. The luminosity has a short-lived spike whose duration, ~ 0.1 Myr, is similar to estimates of the duration of the "Class 0" phase.

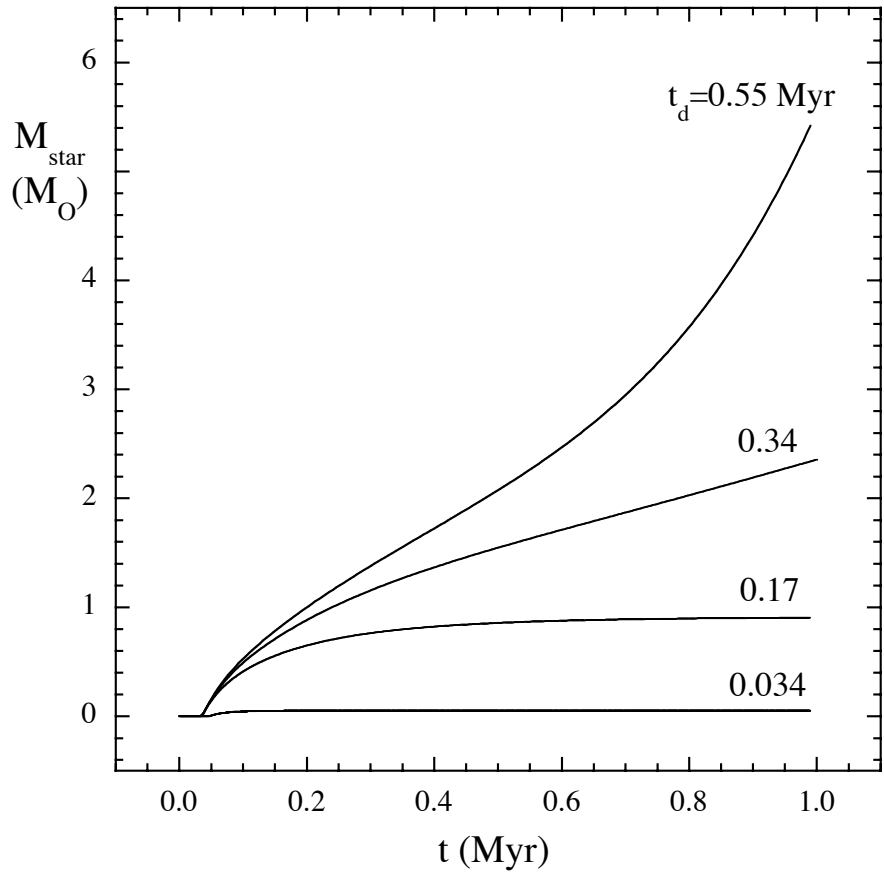

Figure 6. Accretion history for T=10 K, $n_0=10^6$ cm$^{-3}$, $n_u=10^3$ cm$^{-3}$ and for dispersal time scales from 0.034 Myr to 0.55 Myr. The four cases illustrate "self-limiting" accretion (lower two curves) and "runaway" accretion (upper curve) as the dispersal time increases.

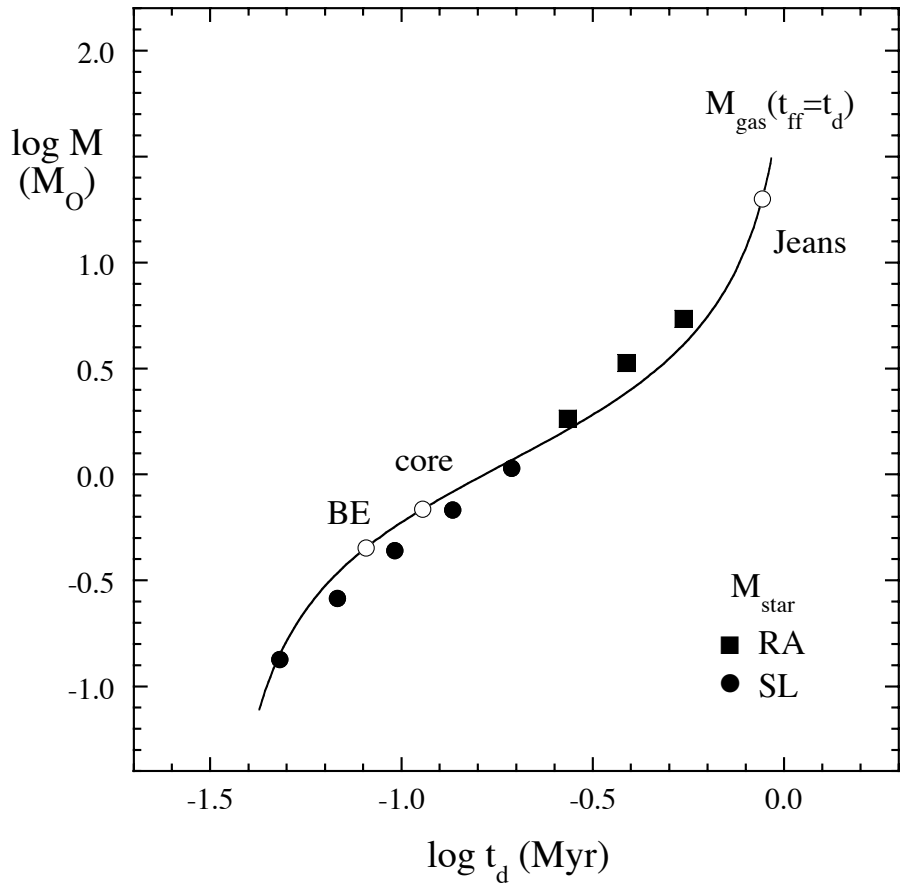

Figure 7. Gas and protostar masses as functions of time. The solid curve shows the gas mass whose free-fall time is $t_d$. The open circles show the critical Bonnor-Ebert mass, the core mass, and the Jeans mass for gas temperature and density as in Figures 1 and 2. The filled symbols indicate final protostar masses arising from "self-limiting" accretion (circles) and "runaway" accretion (squares).

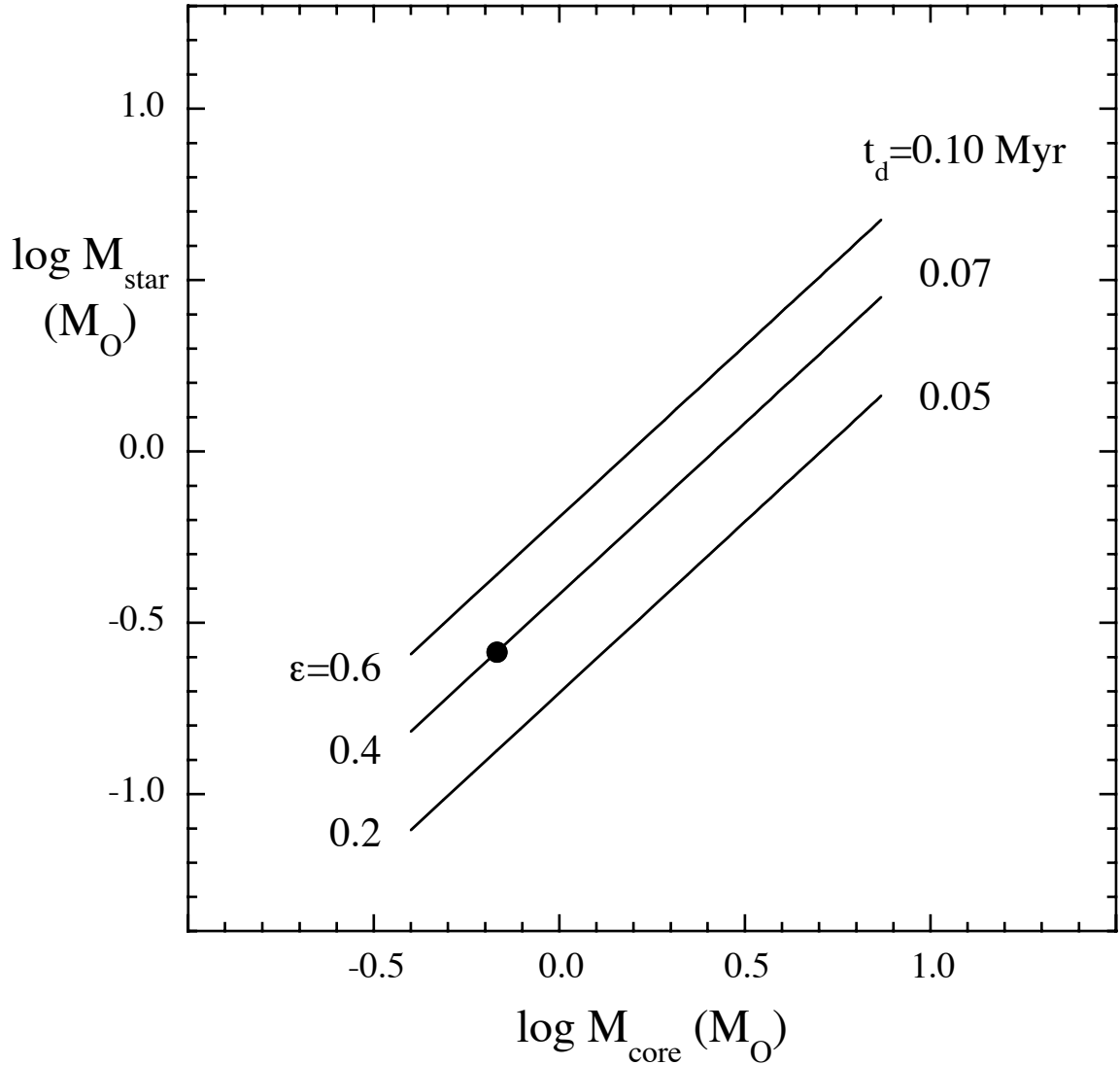

Figure 8. Final self-limiting protostar mass as a function of initial core mass, for peak and background densities $10^6$ and $10^3$ cm$^{-3}$, gas temperature 7-50 K, and three dispersal times $t_d$ as shown. Each curve is labelled with the value $\varepsilon = M_{star}/M_{core}$. The filled circle marks the protostar mass 0.26 $M_O$, near the peak of the IMF.